\documentclass[runningheads]{llncs}

\usepackage[T1]{fontenc}
\usepackage{graphicx}
\usepackage{url}
\usepackage[utf8]{inputenc}
\usepackage[all]{xy}
\usepackage{amsmath}
\usepackage{todonotes}
\usepackage{array}
\usepackage{listings}
\usepackage[a4paper]{geometry}
\usepackage{csquotes}
\usepackage{amsmath,amssymb}
\usepackage{float}%
\usepackage{hyperref}
\begin{document}
\title{The Role of Valence and Meta-awareness in Mirror Self-recognition Using Hierarchical Active Inference}
%
%
\author{Jonathan Bauermeister\inst{1}\and
 Pablo Lanillos\inst{1}}
\authorrunning{J. Bauermeister and P. Lanillos}
\institute{Radboud University, Houtlaan 4, 6525 XZ Nijmegen, NL}
%
\maketitle              
\begin{abstract}
The underlying processes that enable self-perception are crucial for understanding multisensory integration, body perception and action, and the development of the self. Previous computational models have overlooked an essential aspect: affective or emotional components cannot be uncoupled from the self-recognition process. Hence, here we propose a computational approach to study self-recognition that incorporates \textit{affect} using state-of-the-art hierarchical active inference. We evaluated our model in a synthetic experiment inspired by the mirror self-recognition test, a benchmark for evaluating self-recognition in animals and humans alike. Results show that $i$) negative valence arises when the agent recognizes itself and learns something unexpected about its internal states. Furthermore, $ii$) the agent in the presence of strong prior expectations of a negative affective state will avoid the mirror altogether in anticipation of an undesired learning process. Both results are in line with current literature on human self-recognition.
\keywords{Active Inference  \and Affect \and Self-mirror recognition}
\end{abstract}

\section{Introduction}\label{sec:introduction}
The ability of self-recognition has been typically attributed only to humans and few other species~\cite{Anderson2011} and hides several essential brain processes related to multisensory perception, embodiment and decision making~\cite{mci/Hoffmann2021,lanillos2017enactive}. To evaluate this ability, Gallup, in 1970, developed a test for chimpanzees named the mirror self-recognition (MSR)~\cite{gallup1970chimpanzees}. This test, which was also adapted for infants~\cite{Amsterdam1972}, consists of placing a mark, unbeknownst to the subject, on her face. The subject is then placed in front of a mirror. The agent passes the test if there are reaching or exploratory behaviours to remove the mark or inspect it.

While most studies postulate mark directed behaviour (inspect or remove) as a necessary condition for self-recognition \cite{Bard2006}, more recent cross-cultural studies have shown that children from cultures with higher parental authority are not inclined to remove the mark \cite{Broesch2010}. Similarly if one creates a social context during the mirror test, where several other subjects around the infant also have marks on their face, the infant despite having passed the mirror test before is less motivated to engage in mark directed behaviour \cite{Rochat2012}. Both results, by enriching the complexity of such behaviour by environmental and social factors, cast doubts on interpreting the necessity or sufficiency of mark directed behaviour for self-recognition. 

On the other hand, it has been evidenced in humans a strong emotional component, i.e., to express negative affect when seeing their own reflection. When infants pass the test around the age of two, they universally express negative affect toward their mirror image, which has been interpreted as embarrassment, shyness or puzzlement \cite{Amsterdam1972,Rochat2003}. 
Ultimately, we do not know what the phenomenology of a two-year-old seeing herself in the mirror is like. Anyhow, \textit{the negative} 
\textit{affective part of the experience seems to be uncontested}.

While there is previous research on computational models of self-recognition (e.g., generative modelling focusing on visual-kinesthetic matching or appearance cues~\cite{Lanillos2020,mci/Hoffmann2021}), none of the works has attended to the emotional component. Here, we studied self-recognition by ($i$) developing of a computational model that incorporates the affective component into the self-recognition process and ($ii$) evaluating it on a new synthetic experiment based on the MSR.

To model the affective component we use the notion of valence~\cite{Hesp2021}. The working hypothesis is that the negative or positive quality of an affective experience can arise as a consequence of obtaining new information about oneself through mirror self-recognition. Importantly, this new information might favour different action selection policies (action dependent valence), leading to a change in valence. This iterative process coherently connects emotions with self-recognition and decision-making. To incorporate valence into the perception-action loop we used the hierarchical active inference construct~\cite{Friston2017}, where the agent perceives, learns and acts to obtain the expected outcomes by minimizing the expected free energy.

We further developed an experimental benchmark to evaluate the effect of valence in the process of decision-making and self-recognition. In the experiment, the agent can decide to look at a mirror, look at a wall or look at a video of another agent. Furthermore, thanks to the hierarchical nature of the model, we further studied the importance of meta-cognition (`higher' layers), in combination with affect, for (anticipated) self-recognition. For instance, adults in full possession of a self-concept can also anticipate a confrontation with their mirror image. If self-evaluation is negative, or one's body image has radically shifted due to surgery, patients are motivated to actively avoid the mirror~\cite{Freysteinson2012}.\\

\noindent\textbf{Related Work}\label{sec:relatedwork}. 
There were several tries to build a computational model of self-recognition---See \cite{mci/Hoffmann2021} for a review. Relevant to this work, in \cite{Lanillos2020} the robot inferred itself by answering the question 'did I generate those sensory outcomes?'. For example, if the robot has an intention to move its arm and can predict its interoceptive and exteroceptive sensory outcomes with low prediction error, then it will infer that the likeliest cause of this action was the system itself. While this approach may be promising to give insights into self-recognition, it does not yet explain how affect arises during the human MSR test. 
It has been theorized that affective and action based self-modelling naturally arises for a system engaged in deep temporal active inference \cite{Deane2020,Deane2021}. Nevertheless, to the authors knowledge, there is no computational model as of now, that explicitly assesses affect within self-recognition.
Fortunately, recently within the active inference research, there has been an effort to introduce internal drivers that modulate the generative model parameters and the action selection process. For instance, valence -- pleasantness or unpleasantness of an emotional stimulus -- was introduced in \cite{Hesp2021} to model the confidence of the model estimates. Importantly, valence encodes how well the agent is performing in the environment, thus, aiding action selection. The mathematical formalization of valence and the hierarchical structure of the generative model allows the building of a complex agent that has beliefs over beliefs, which are modulated by the increase or decrease of valence, thus affecting the whole decision making process. In this work, we adapt this model to study self-recognition and decision making.



\section{Methods}\label{sec:methods}
First, we describe the general framework of our approach and how to introduce valence based on the work of~\cite{Hesp2021}. Second, we detail the affective self-recognition computational model and finally, we describe the experimental setup for self-recognition.
\\

\subsection{Discrete hierarchical active inference} 
We model the problem under the discrete state-space formulation of active inference~\cite{Parr2019}\footnote{For a thorough tutorial on discrete active inference formulation see~\cite{Smith2021a} and  for a concise mathematical overview see~\cite{Costa2020}.}. The agent computes both the posterior state estimation (perception) and action selection by minimizing, through marginal message passing, a single quantity: the expected free energy. This quantity measures the divergence from the current expectations to the real world state. In order to be able to compute it the agent needs a generative model of the world, thus, allowing predictions of the future outcomes. See Appendix~\ref{sec:appendix} for detailed explanation.
\\

\begin{figure}[h]
    \centering
    \includegraphics[width = 1 \textwidth]{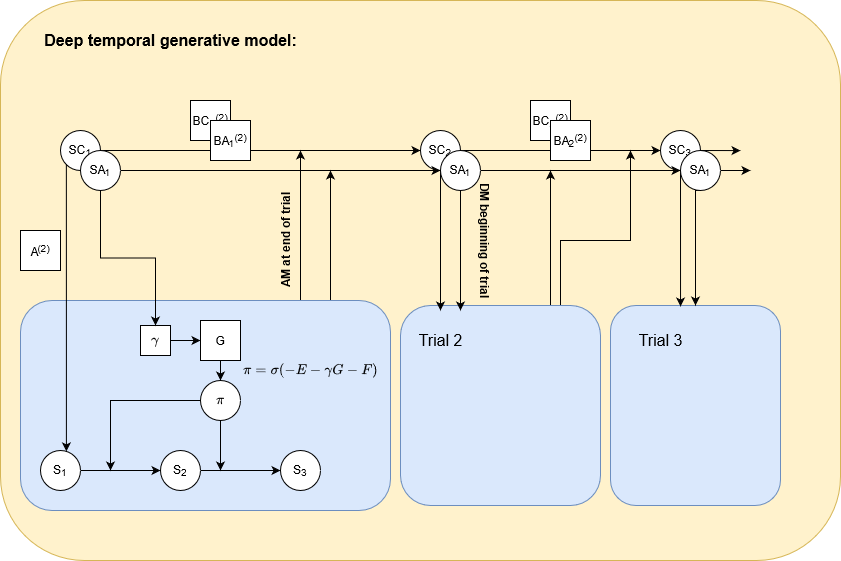}
    \caption{A temporally deep generative model with two hierarchical layers. The blue boxes in the first layer correspond to trials (here simplified). This is the architecture the agent uses to perceive and act in the world. The second layer only communicates with the first at the beginning of a trial through descending Messages and at the end of the trial is informed by ascending messages. Hence, it can already be seen that states on the second level change slower (only once every trial) than states on the first layer. Here the agent has two state factors on the second level. It has a contextual belief which replaces the prior at the beginning of each trial. Additionally, it has an affective state which sets the precision on G at the beginning of the trial. The impact of G on $\pi$ can now be regulated by the agent through its affective state. Figure adapted from \cite{Hesp2021}.}
    \label{fig:twoLayers}
\end{figure}

\noindent\textbf{Temporal Depth}\label{sec:temporaldepth}
To achieve temporal depth we use a hierarchical generative model as depicted in Fig. \ref{fig:twoLayers}. Here, the hidden states on the second layer change slower than the hidden states on the first layer. Thus the beliefs about states in the first layer (bottom) can fluctuate several times within one trial, where the beliefs on the second layer (top) only change at the end of each trial (the length of a trial is defined by the modeller). For example, the agent can have the abstract belief that it is in a happy mood. This belief will set the priors on the first level at the beginning of a trial accordingly. So now the agent expects certain observations (facial muscles expressing a smile, heart rate going up etc.), even if within the trial the facial muscles will most likely change several times (depending on the granularity of the model) and not only stay in one position, say smiling, the agent could still infer that overall it is happy (which is an abstract second-order belief that integrates information over time). Only if in the course of the trial it consistently observes unexpected observations (prediction errors) it will update its belief on the second level at the end of a trial accordingly.\\
Mathematically, the second layer has likelihood mappings and state transitions like the first layer. They differ in that the likelihood mapping A2 does not map from observation to hidden state but from the hidden state at layer one (facial expression i.e smile, frown, neutral) to the hidden state on layer two (mood i.e happy or sad). The transition matrix B2 then encodes how likely the context, for example, the agent's mood, changes over trials. As a consequence, the prior D1 is replaced by a more dynamically changing higher-order state. At the beginning of each trial, the higher-order state acts as prior for the agent. By the end of the trial the agent updates the higher-order state based on the information collected in this trial. For a mathematical expression of ascending and descending messages---See~\cite{Hesp2021} for further description.\\

\noindent\textbf{Meta-cognition and Valence.} An agent equipped with such a deep temporal model can learn context and perform some tasks very well but underperforms in a volatile changing environment. 
Here we describe, based on the work of~\cite{Hesp2021}, how affect can be included in a discrete hierarchical active inference network. Affect can be formalized through valence (negative or positive). Valence can be explained as the expression of confidence in the model estimates. If the agent's actions continuously lead to the outcomes that it expects and prefers, it grows more confident in its action model and weighs it stronger as acquired habits. Whereas if the environment is very volatile and it cannot rely on its learned action model yielding to an 'anxious' state. 
The agent equipped with affective states finds better and biologically more plausible action plans (policies) than one without~\cite{Hesp2021}. Partly because it takes time to construct a reliable action model that tells the agent which policies to take under which circumstances. Hence, when the environment changes fast and unexpected the action model (learn by experience) might become completely useless. An affective agent that reacts with negative valence towards the unexpected change in the environment will able to quickly adapt by lowering the precision of its action model to reevaluate the new situation. Conversely, an agent without affective states cannot quickly adapt and will execute the same actions despite an environment that has changed.

Thus, valence acts as a second-order state. Implementation-wise the agent has a categorical distribution over it being either in the state 'positive valence' or 'negative valence', and it is updated at the end of a trial via ascending messages. If in a trial the agent could rely on its action model then it increases its valence for the next trial. At the beginning of the next trial, instead of using a static prior the second-order affective state informs the \textit{precision} of the action model. Such a top-down estimation of the reliability of its model can also be understood as a form of meta-cognition as it is monitoring the confidence of the cognitive process.\\


\noindent \textbf{Meta-awareness and attention.} The meta-cognition architecture can be extended to model meta-awareness by adding a third layer, and allowing the agent to dynamically change the information flow between second and first layer. This has been implemented by~\cite{SandvedSmith2021} as a precision on the likelihood mapping. Generally, precision on likelihood matrices in active inference is linked to attentional processes. The states on the third level then represent if the agent is aware of her cognitive processes on the second level. The benefits and biological plausibility of this meta-awareness capacity have been shown by \cite{SandvedSmith2021}. 
Here we use a simplified setup where we change the precision of the 2nd layer likelihood by hand to see how the agent's behaviour changes if it is very aware of its emotional states and hence, can use the information for cognition deeper down the hierarchy. The precision modulates how strong the connection between the two layers is and therefore how informative descending and ascending messages are. For instance, low precision will lead the agent to rely more on its `direct experience' (first layer of its generative model).

\subsection{Affective self-recognition model}\label{sec:affectivemodel}
Based on the previously described framework, we propose a computational model to investigate self-recognition with the emotional component. We focused on the following two research questions: ($i$) How does valence influence behaviour during mirror self-recognition? and ($ii$) How might negative valence arise in mirror self-recognition?

To evaluate the model and further be able to answer the questions, we designed an experiment where an agent can decide to either see its emotional expression in the mirror, look at a wall and see no face or look at a video of an emotional expression of another person. Note that our computer-simulated agent can not look into an actual physical mirror, so we need to formalize the function of the mirror, which possibly leads to an affective reaction. 
We interpret making an observation in the mirror as acquiring information about oneself (here: about the agent's emotional state via its facial expression) by dragging the internal attention of the agent onto this aspect of itself. Thus, the mirror is a self-exploration tool that allows this information to be available for decision making and introspection from layers higher up the hierarchy.\\

\begin{figure}[hbtp!]
    \centering
    \includegraphics[width=0.7\textwidth, height=145px]{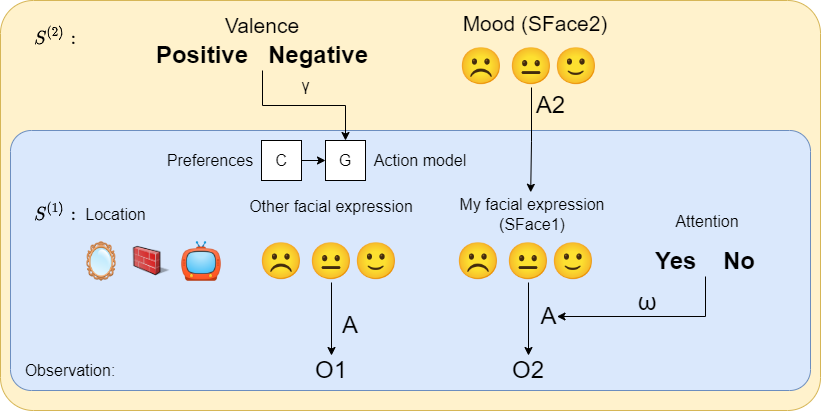}
    \caption{\textbf{Generative model description}. Two layered model of an agent inferring its own mood and deciding weather to look at the mirror or not. At any given time step the 1st layer includes an action model G, preferences C and four state factors: Location, Other-facial-expression, My-facial-expression and Attention. The Attention state modulates the likelihood A via the precision $\omega$. For each state factor, the agent has a categorical distribution of which state it believes itself to be in. The 2nd layer tracks the agent's valence and belief about its mood. Valence interacts with the action model G and the mood or context state interacts with the agents belief of its facial expression on the first level via A2.}
    \label{generative}
\end{figure}

\noindent \textbf{Generative Model.} We formalize a two-layered deep generative model to capture the self-recognition experiment, as described in Fig. \ref{generative}.
The agent can obtain exteroceptive observations, where it either sees a face that is happy, neutral or sad, or nothing when it looks at the wall. This information can be used to infer the emotional state of the other person. And it can obtain interoceptive observations about its facial expression (for example sensing its facial muscles) to infer its emotional state (happy, neutral, sad). The agent also has a state (attention in the figure) that captures if it is paying attention to its interoceptive observation with values Yes or No. This is an attention state, in the sense that it modulates the precision on the likelihood A, but it cannot be actively controlled by the agent. Instead, it captures what happens internally when the agent sees itself in the mirror. By recognizing itself, it is forced to pay attention to its internal observation (the precision $\omega$ on A will become very precise). Hence, formalizing the notion of the mirror dragging attention onto oneself and making certain observations informative.

On the second layer, the agent has two state factors. The valence can be positive or negative (implemented as a categorical distribution). Its value depends on the expected precision of the action model G. Additionally, the agent has a second-order belief about which mood it is in (happy, neutral, sad). This state sets the priors of what facial expression the agent expects when observing itself. For all the details of how this generative model is implemented please see the Appendix~\ref{sec:appendix}.\\

\noindent\textbf{Agent operational specification.}
\label{sec:experimentalsetup}
 First, inspired by the idea that if self-evaluation is positive one seeks out a mirror, we assume that the agent prefers to see \textit{itself} but only if it is 'happy' or `neutral' instead of `sad'. This is encoded in the preference matrix C. The agent's actions are to go to one of the three locations: mirror, wall or tv. Because the agent knows from its generative model that it will pay attention to itself if it goes to the mirror its behaviour will depend on its self-knowledge, i.e., what state it believes to be in and how aware it is of that state. If it thinks it is happy one would expect that the agent will act to admire itself in the mirror. Second, the true emotional state of the agent may change. Here we have coupled the dynamics of the true state to the current valence of the agent. If its positive valence goes above 70 \% its true state shifts to happy, below 30\% to sad and otherwise neutral. To be robust against small fluctuations, we imposed that the agent's belief about its valence has to shift at least by 15\% to make a switch. Although, these decisions are arbitrary, they are designed to show how the agent adapts to a change in its true state.

\section{Results}\label{sec:results}
we analyzed our model behaviour in our synthetic experiment to study which actions the agent chooses and when and how the valence of the agent changes under different conditions, such as changing the true emotional state, the prior knowledge the agent has about its emotional state or the introspective availability of its emotional states (precision on A2). Particularly, we focused on two different initial conditions. In the first experiment (Sec.~\ref{sec:cond1}), we study how valence might naturally evolve in a mirror self-recognition scenario. To this end, the agents true state was set to sad, but it had low meta-awareness. For the second experiment (Sec.~\ref{sec:cond2}), we studied mirror avoidance behaviour. Thus, the agent had high meta-awareness. The code to replicate the results can be found in this \href{https://github.com/blindreview}{this link}. 

Each agent was evaluated for 8 consecutive trials in each condition, where each trial lasts for three time steps (three observations). After the first observation, the agent will have to decide where to go (mirror, wall or tv). Its action plan horizon is two steps ahead, thus, it can predict outcomes until the end of a trial by using its generative model. 

\subsection{Experiment 1: I am sad and I know it, but I am not very aware of it}
\label{sec:cond1}
\begin{figure}[h]
    \centering
    \includegraphics[width = 1.0 \textwidth]{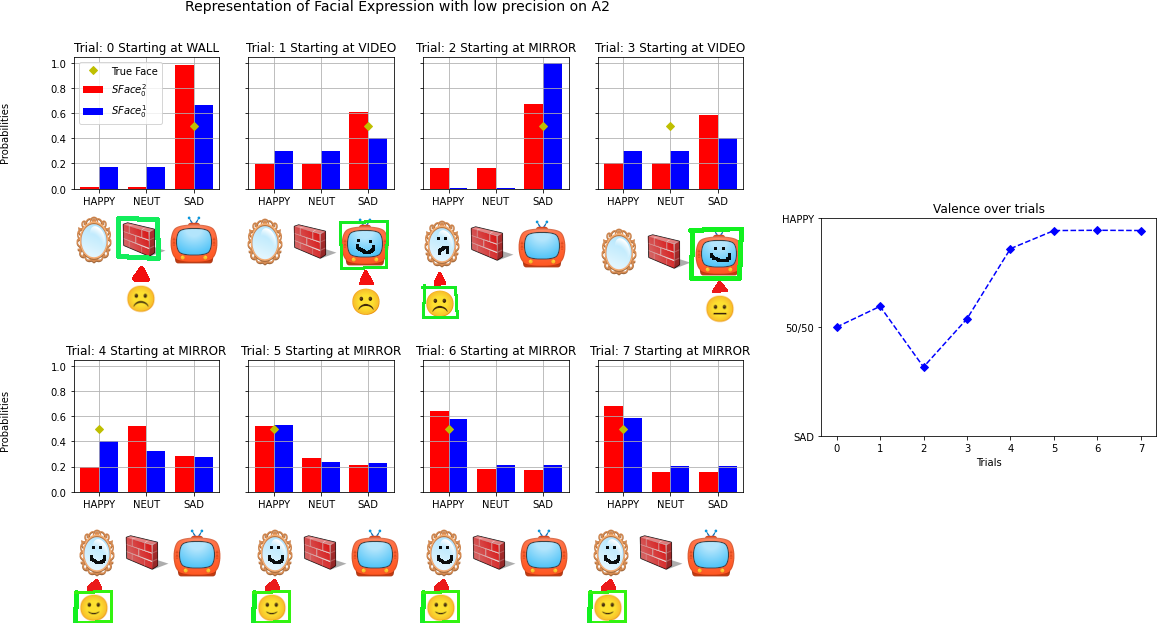}
    \caption{\textbf{Experiment 1}. The belief distribution of My- facial-expression (SFace1 and SFace2) at the beginning of each trial are shown. Below that the agent with its true state is shown as smiley. It is indicated at what location it currently is. The green box indicates if its attention is on the exteroceptive or interoceptive observation. The graph next to it shows how the agent's valence evolves throughout the trials. The value is a probability, where a high value means high confidence in being in the state of `positive valence'. At each trial, the agent has to choose where to go and hence at which location it will start the next trial. The valence graph shows how negative affect is elicited when the agent makes an unexpected observation in the mirror and therefore learns something new about its internal states. 
    }
    \label{result1}
\end{figure}

We set the true state of the agent to `sad' and the precision on A2 low. Also, the agent `knows' that it is sad on the second level of the hierarchy. However, due to the low precision on A2 this will not inform the first level, thus, the agent will be more informed by the actual perceptual information in a given trial. 
The experiment is described in Fig.~\ref{result1}. At trial 0 the agent is convinced enough that it is sad and calculates that its best action will be to go to the video. Paying attention to the other face, loosens its priors making them less informative about its emotional state. This is reflected in the categorical distribution at trial 1. It is more entropic or less precise as in trial 0. In trial 1 it decides to go to the mirror. Hence in trial 2, it makes an unexpected observation (seeing itself frown in the mirror) which also results in a drop in valence, indicating that this particular mirror encounter is negatively experienced. Having reaffirmed its belief that it is sad, it finds it best to go to the video. With this decision, the agent regains a bit of confidence in its action model. The valence goes up between trials 2 and 3. And due to the in build dynamics, its true state shifts to neutral.  
 Finally,  the agent notices the change in its true state to neutral and decides it's time to go to the mirror again, which even further improves its confidence in its action model and for the rest of the trials it will be happily smiling at itself in the mirror.

\subsection{Experiment 2: I am sad and I know it and I am aware of it}
\label{sec:cond2}
\begin{figure}[h!]
    \centering
    \includegraphics[width = 1 \textwidth]{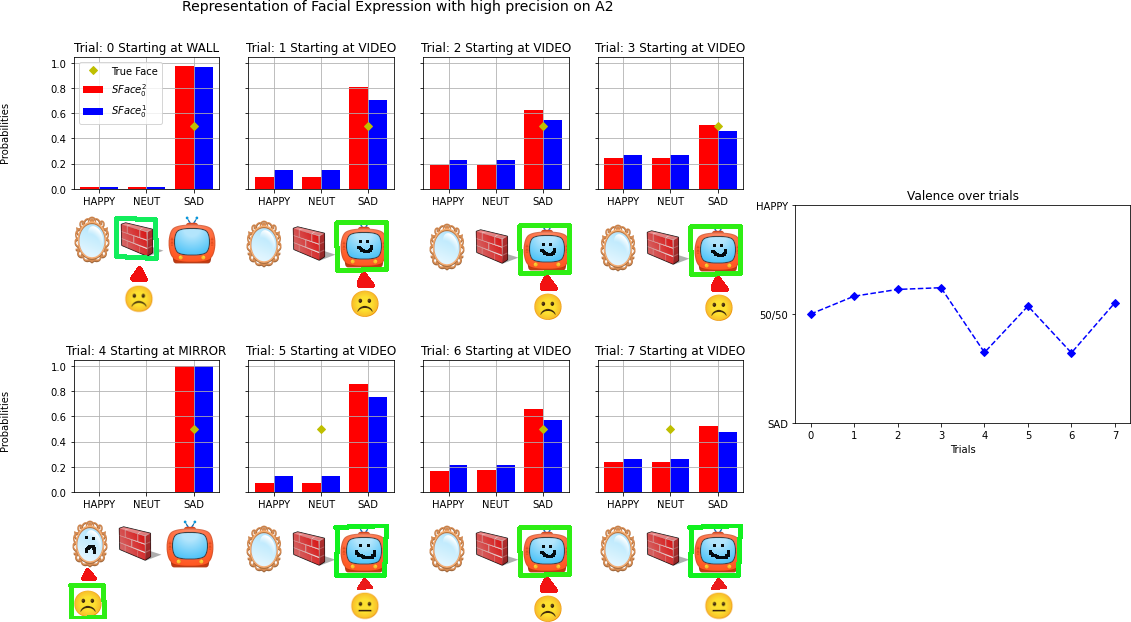}
    \caption{\textbf{Experiment 2.} The agent, now being more aware of its own internal state, is anticipating an uncanny encounter with the mirror. Hence it is avoiding the mirror longer. However, it is also less able to pick up on a change in the true state of its emotion due to it expecting, much stronger, that it is actually sad. Its valence only shifts between sad and neutral. The results show how having strong priors about its emotional state and being able to attend to it discourages exploratory behaviour such as going to the mirror and learning about itself.}
    \label{result2}
\end{figure}

We explored how the behaviour of the agent changes if its first-order states are introspectively available to it. The experiment is described in Fig.~\ref{result2}. We set the same initial state as in the previous study: true state is `sad' and first location is the `wall'. Differently, this time the mapping A2 is very precise.

At the first time point of each trial, the beliefs on both levels of the hierarchy are the same due to the almost one-to-one mapping of A2. The agent's behaviour differs from previous experiment in that it decides to stay at the video until trial 3. Only after a much longer time---when its priors have loosened enough---it tries out the mirror again. When this happens (in trial 4) we observe again a reconfirmation of its belief of being sad. This is accompanied by the same drop in valence once it realizes it wasn't the best action to go to the mirror. When going back to the video the agent has the chance to pick up on the change of its true state. However, it misses it because it keeps its prior belief of being sad extended through time.

\section{Discussion}

The experiments showed a possible self-recognition process where the agent gets insight into its own generative model due to observations about itself made available through the mirror. The valence of the agent was coupled to this observation being surprising or not. The results show, how the valence changes and how the agents favoured actions change as a result of a change in valence. Thus, modelling mirror self-recognition as an internal shift of attention shows how negative and positive valence plausibly arises. The mirror self-recognition provides the agent with new self-knowledge, which can be used by deeper levels in the hierarchy to perform further inference. For example, changing precision estimates, thereby possibly favouring different actions, which in turn results in a change in action-based valence. We do not model how self-knowledge first arises, but what can be shown here is that negative valence arises in self-recognition processes that yield insights about oneself, which change the best available action. The negative valence is not directly dependent on the agent feeling sad or happy (its emotional states that the agents tries to infer), but rather about the (accurate) knowledge the agent has about these states. It is important to highlight that emotion is a more complex phenomenon that is likely constituted by many more dimensions than just valence \cite{Fontaine2007}. Therefore, this computational model only offers the first steps, namely trying to account for the valenced part of mirror self-recognition. Besides, only using a categorical attention internal state is a strong simplification of the reflection of one's physical appearance for visual-kinesthetic matching as shown by \cite{Lanillos2020,mci/Hoffmann2021}. Mirror self-recognition in humans may additionally involve further internal attentional dynamics.\\
 
\noindent \textbf{Meta-awareness.} The capacity of meta-awareness allows an agent to change the strength with which one is aware of oneself. From dreaming to being awake, from being lost in thought to paying attention, humans in full possession of a self-concept do it all the time. The model behaviour in experiment 2 (Fig.~\ref{result2}) shows how meta-awareness is important to explain mirror avoidance and engagement behaviour. Being highly aware of a negative state of self an agent can anticipate an unsettling mirror encounter and prefers to avoid the mirror. Although at the cost of potentially missing a change in its true state. Given the limitations of the model, these statements are speculative. By expanding the model in future research one can potentially address open questions such as mirror avoidance and modelling mirrors in therapy~\cite{Freysteinson2020}. Airing on the side of caution, even if the proposed computational model here does not simulate self-awareness, it can be used to pose interesting questions about action dependent affect in mirror self-recognition for future work.
For example what are the actions available to an infant recognizing itself in the mirror? Is its negative affect resulting from suddenly being suspect of its usual policy of playful engagement with the other in the mirror? Or is it a feeling of alienation? If one prefers to interpret the negative affect as a feeling of alienation one could argue to expand the model to include mental actions. Planning on the second level (mental actions) could have its own confidence and valence associated with them. Actions on this (or even higher levels) could answer more existential questions such as what kind of person should I be? How do others see me? Tracking the expected confidence in one's mental actions might be an interesting choice to model more complex emotions such as the feeling of alienation. It could be interesting to design clever mirror tests, that involve different action affordances to test different stages of self-awareness more specifically. 

\section{Conclusion}
This thesis proposes an affective self-recognition model based on the formalization of action dependent valence, using hierarchical active inference. As a proof of concept, we have shown how a synthetic affective response towards one's mirror image might arise. The results show that mirror self-recognition provides the agent with new information, which changes the favoured strategy and hence leads to negative valence. Secondly, the results show how an active inference agent with high meta-awareness of a negative evaluated state of self displays mirror avoidance behaviour. Therefore emphasizing the importance of deeper hierarchical layers, regarded as meta-cognition and meta-awareness, to explain more complex behaviours seen when facing the MSR test. 

\bibliographystyle{splncs04}
\bibliography{IWAI2022-affective}

\appendix
\newpage
\section*{Appendix}
\label{sec:appendix}
\section*{Discrete Hierarchical Active Inference}
\begin{figure}[hbtp!]
    \centering
    \includegraphics[width=1.3\textwidth]{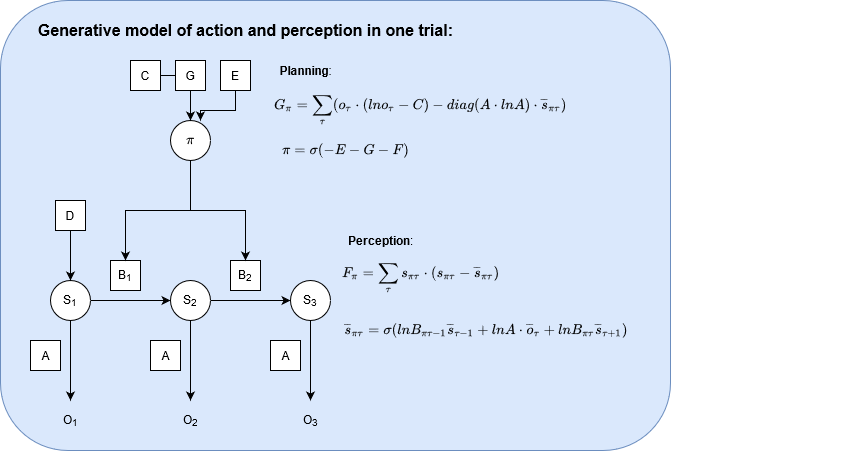}
    \caption{A generative model of one trial with three time steps. Rectangles correspond to categorical distributions and circles to the random variables the agent wants to learn (here hidden states and policies). A is the likelihood that defines how likely observations are given the state $P(o_{\tau} \vert s_{\tau})$. B encodes the probability of moving from one state into the next one $P(s_{\tau+1} \vert s_{\tau}, \pi)$ given the policy $\pi$. C is a vector or matrix that encodes which observations the agent prefers. D gives the prior at the first time step in the trial to perform Bayesian inference. G is the expected free energy. The best policy is the one that minimizes G (future reward + information gain) and F (current perceptual evidence or prediction error). F is also calculated for each policy meaning that the agent has a posterior state estimate for all possible policies. Lastly, E sets a `habitual' prior for policies in case G is uninformative.  Note that the past message $\ln B_{\pi\tau-1}\overline{s}_{\tau-1}$ at the first time point becomes the prior $\ln D$ and at the last time point the future message becomes ones (hence uninformative). Finally, the actual observation is marked with a bar $\overline{o}$ in contrast to the predictive posterior over observations $o$. Figure adapted from \cite{Hesp2021}.}
    \label{fig:generativemodel}
\end{figure}

Figure \ref{fig:generativemodel} shows a generative model used for discrete hierarchical active inference. The agent's hidden state is $s_\tau$, where $\tau$ indexes the time step. At each step, the agent gets an observation from the environment $o_\tau$. Following active inference simplified notation~\cite{Costa2020} we will use capital letters to define the probabilistic functions. The agent has a prior belief D about hidden states $s_\tau$, a likelihood mapping A between states and observations ($P(o_\tau|s_\tau)$), and a transitions matrix B that encodes how states evolve over time depending on the policy $\pi$ ($P(s_{\tau+1} \vert s_{\tau}, \pi)$). The agent can invert the generative model to perform Bayesian inference and get from an observation to a posterior over hidden states (in active inference this inference process is equated with perception). \\
\\
To encode the intention or goal, the agent has preferred observations defined by the matrix C. By minimizing the expected free energy the agent chooses a policy that changes the hidden states such that they are likely to produce preferred observations (and minimize overall perceptual ambiguity). The action model G uses those preferences to track how well each policy $\pi$ is expected to achieve this goal.\\
\\
To understand the computations we will describe an agent that performs a trial with three time steps, meaning it has three observations $\tau =\{1,2,3\}$, as described in Fig.~\ref{fig:generativemodel}. Here an actual observation is denoted with a bar $\overline{o}$ in contrast to the probability distribution of expected observations $o$. From the first observation the agent infers the posterior hidden state $\overline{s}_\tau$ at time instant $\tau = 0$ through Bayesian inference, via the likelihood matrix A\footnote{Using the discrete space formulation of active inference in matrix form this is computed by selecting the right column of the matrix A, i.e., through a one-hot observational vector.} and the prior D:
\begin{center}
  $\overline{s}_{\tau} = \ln A \cdot \overline{o}_{\tau} + \ln D$   
\end{center}
Note that we almost get classical Bayesian inference (likelihood multiplied by the prior), but without normalizing by the evidence term (the multiplication turns into addition due to working in logarithmic space). The evidence term is mostly intractable in larger models. So to compute the full posterior, we can alternatively minimize the variational free energy bound instead~\cite{Friston2017}. This free energy formulation in discrete state space boils down to the difference in the belief the agent has about the world before (prior $s_{\tau}$) and after (posterior $\overline{s}_{\tau}$) an observation.
\begin{center}
  $F = s_{\tau} - \overline{s}_{\tau}$  
\end{center}
In other words, the free energy encodes the prediction error. If the prior belief matches or is supported by the observation the free energy is low. In contrast to a surprising observation that renders the prior belief less likely and therefore increases F. 
The agent can predict observations with the generative model and the belief about hidden states. It evaluates these predictions by how well they compare to the actual evidence, by calculating F. Then the agent can iteratively make predictions that will decrease F and hence lead to more accurate estimates of hidden states. \\
\\
We have described how the agent updates its belief $\overline{s}_\tau$ about the world by trying to minimize prediction errors, therefore getting good at expecting what is really out there. Next, the agent also computes the expected observations to optimize not only for the current time point but for the whole trial. Under each policy or plan of actions $\pi$ the agent can evaluate, how likely certain observations are in the future. Additionally, it can consider how ambiguous possible future observations are. Both, information gain and preferred observations, are described in the expected free energy G:
\begin{center}
  $G = \sum_{\tau} (o_{\tau} \cdot (\ln o_{\tau} - C) - \text{diag}(A \cdot \ln A) \cdot \overline{s}_{\tau})$  
\end{center}
The first part of the equation is the average difference in expected observations $o_\tau$ and preferred outcomes C over all time points. The second part relates to the model entropy or how precise the distribution is from which the expected observations are sampled. For each state at time $\tau$ there is a likelihood A that can give the agent more or less certainty about what outcome to expect.

To sum up, the agent minimizes $F$ to optimize the posterior belief about states (estimation) and minimizes G to compute which policy to choose (action). Lastly, marginal message passing is just a mathematical way of sending information across time. For example, if the agent already knows which policy it is likely to take after seeing the first observation then that knowledge can, through marginal message passing, already inform its prior at the next time point in the trial. Vice versa the agent can update past beliefs, based on new observations which later can be helpful for learning. This leaves us with the final equations for posterior state estimation including future and past messages (using the transition matrix B) and the average free energy over timepoints in one trial: 
\begin{center}
\begin{align}
        \overline{s}_{\tau} &= \sigma(\ln B_{\tau-1} \overline{s}_{\tau-1} + \ln A \cdot \overline{o}_{\tau} + \ln B_{\tau}\overline{s}_{\tau+1})  \\
    F &= \sum_{\tau} s_{\tau} \cdot (s_{\tau} - \overline{s}_{\tau})
\end{align}
\end{center}
Where $\sigma$ is a softmax function that normalizes the input vector such that it sums to 1 and forms a proper probability distribution.
The G and these two equations defined as shown in Fig. \ref{fig:generativemodel}, describe the agents' basis to act in the world within a given trial. A limitation of this scheme is the static nature of the prior D at the beginning of each trial. It would be preferable that the agent can update/learn its prior based on the information it gathered in a trial. To make the context in which the agent navigates learnable one can expand the generative model with a deep temporal layer~\cite{Friston2017}. This allows the agent to form abstract and contextual beliefs that carry across trials---as described in the next subsection.

\section*{Affective self-recognition model implementation details}
This section provides details about the generative model. Simulations were run by extending the pymdp infer-actively framework on \href{https://github.com/infer-actively/pymdp}{github}. The inference process of state estimation and policy selection on the first layer has been calculated using the pymdp framework. Inference on the second level, via ascending and descending messages was programmed for this setup. A commented code is available on github via \href{https://github.com/blindreview}{this link}. 

\subsection*{First Layer}
The priors on the state factors are specified in the D matrix. For the state factor 'Location' (Mirror, Wall, Video) the prior is uniform. The state factor 'Other emotional state' (Happy, Neutral, Sad, Null), which can be inferred via the observations (Smile, Neutral, Frown, None) also has a uniform prior. The prior on 'Self emotional state' depends on the starting condition and the second layer. Lastly, the state 'Mirror-controlled attention' (don't attend, attend) is set on don't attend:
\\ \\
$P(S^{MC-Attention}_{\tau_0}) = [\textbf{0.99, 0.01}]$
\\ \\
    For each observation, there is a likelihood tensor $A_{1-3}$. The first observation is exteroceptive (Smile, Neutral, Frown, None), the second interoceptive (Smile, Neutral, Frown) and the third an observation about the location (which ensures the agent always knows where she is). The dimensions of the likelihoods are the observation and all the hidden state factors, i.e: A[Observation, Location, Other, Self, Attention] or $A_1[4,3,3,3,2]$. 
    For example, if I want to index the likelihood of my exteroceptive observation given that I am looking at the wall:
    \\
    \\
    $P(O_{ex} | S^{Location} = Wall,  S^{Self}, S^{Other}, S^{MC-Attention}) =  \\
    \text{for i,j in 0:2, k in 0:1} \\
     A_{1}[:,1,i,j,k] = \left(\begin{array}{l} \textbf{0.01}  \textit{ Smile} \\ \textbf{0.01} \textit{ Neutral}\\ \textbf{0.01} \textit{ Frown} \\ \textbf{0.97} \textit{ None}\end{array}\right)$
\\
\\
\\
    Basically saying the agent knows her probability of seeing 'None' if she is at the wall is 0.97, independent of all the other states she is in.
    If the agent is in the 'attend' state she is attending to herself and therefore can only relate the information of the exteroceptive observation to herself. This has to be defined for all states, but effectively the agent only makes use of this attention when she is in front of the mirror and the exteroceptive observation in fact relates to her: 
    
    $P(O_{ex} | S^{MC-Attention} = \text{attend},  S^{Self}, S^{Location}) = \\
    \text{for l,i in 0:3 :}\\
    A_{1}[:,l,i,\textbf{:},0] = \left(\begin{array}{l} \textbf{0.97 0.01 0.01}  \textit{ Smile  } \\ \textbf{0.01 0.97 0.01} \textit{ Neutral}\\ \textbf{0.01 0.01 0.97} \textit{ Frown  } \\ \textbf{0.01 0.01 0.01} \textit{ None}\end{array}\right) $
    \\
    \\
    Here the columns stand for the different states in the state factor 'Self emotional state' (Happy, Neutral, Sad). If the agent is not paying attention we get the same matrix, but this time relating to the state of the other.
       \\\\
    $P(O_{ex} | S^{MC-Attention} = \text{don't attend},  S^{Location}, S^{Other}) = \\
    \text{for l,j in 0:3 :}\\
    A_{1}[:,l,\textbf{:},j,1] = \left(\begin{array}{l} \textbf{0.97 0.01 0.01}  \textit{ Smile} \\ \textbf{0.01 0.97 0.01} \textit{ Neutral}\\ \textbf{0.01 0.01 0.97} \textit{ Frown} \\ \textbf{0.01 0.01 0.01} \textit{ None}\end{array}\right) $
    \\\\
    
    Now for the interoceptive observation, the precision on A will depend on the state of attention the agent is in. Therefore one can push A through a softmax with a precision (inverse temperature) parameter c.
    \\ \\
    $P(O_{in} | S^{MC-Attention}, S^{Location}, S^{Other}) = \\
    \text{for l,i in 0:3 and k in 0:1 :}\\
    A_{2}[:,l,i,\textbf{:},0] = \left(\begin{array}{l} \textbf{0.97 0.01 0.01, c}  \textit{ Smile} \\ \textbf{0.01 0.97 0.01, c} \textit{ Neutral}\\ \textbf{0.01 0.01 0.97, c} \textit{ Frown} \end{array}\right) $
    \\\\
    
    Where paying attention has $c = 5$ and not paying attention $c = 0.001$.
    Finally, the location observation is a 1 to 1 mapping:
    \\\\
    $P(O_{loc} | S^{MC-Attention}, S^{Self}, S^{Other} ) = \\
    A_{3}[:,l,i,\textbf{:},0] = \left(\begin{array}{l} \textbf{1 0 0}  \textit{ Mirror} \\ \textbf{0 1 0} \textit{ Wall}\\ \textbf{0 0 1} \textit{ Video} \end{array}\right) $
    \\\\
    Next the transition matrices B need to be defined. The rows correspond to the state in the next time step and columns the state in the current time step. The transition for the location depends on the action chosen and the agent knows with certainty where she will be next. The agent also knows that her attention state will shift to focused when she goes to the mirror and unfocused going to the video. The agent has a bit of uncertainty around how her own emotional state is changing in time and a bit more uncertainty about how the state of the other is changing.
    \\\\
    $P(S^{Self}_{\tau + 1} | S^{Self}_{\tau}) = \\
    B_{1}[:,:,0] = \left(\begin{array}{l} \textbf{0.95 0.05 0.05}  \\ \textbf{0.05 0.95 0.05}\\ \textbf{0.05 0.05 0.95} \end{array}\right)$
    
    \vspace{5ex}
    
    $P(S^{Other}_{\tau + 1} | S^{Other}_{\tau}) = \\
    B_{2}[:,:,0] = \left(\begin{array}{l} \textbf{0.8 0.1 0.1}  \\ \textbf{0.1 0.8 0.1}\\ \textbf{0.1 0.1 0.8} \end{array}\right)$
    \\\\
    
    The preference are set with the C matrix. For all observation modalities C will be initiated with zeros. Then the preference to see self happy or neutral can be encoded as:\\\\
    $C_{1}[0] = \textbf{3.0} \\
    C_{1}[1] = \textbf{3.0}$
    \\\\
    The description of the first layer concludes with the policies available to the agent. They are any combination of going to a location that is possible within a trial. The trials consist of three observations and 2 actions. The agent starts by sampling an observation then decides where to go, and repeats this step. After the final observation, the agent doesn't need to go anywhere because the trial is over and will start again from the beginning.
    
    \section*{Second Layer}
    
    The A and B matrix for the Valence state are the same as in \cite{Hesp2021}:\\\\
    $A2_{valence}[:,:] = \left(\begin{array}{l} \textbf{0.97 0.3}  \\ \textbf{0.3 0.97}\\\end{array}\right)$
    \\
    \vspace{5ex}
    \\
       $B2_{valence}[:,:] = \left(\begin{array}{l} \textbf{0.8 0.3}  \\ \textbf{0.2 0.7}\\\end{array}\right)$
    \\\\
    
    For the state S2Face or 'Mood', the A2 matrix can again be changed with a precision parameter c. This one is set manually to simulate meta-awareness. In my simulation high means c = 5 and low c = 1.
    \\\\
     $A2_{Face}[:,:] = \left(\begin{array}{l} \textbf{1 0, c}  \\ \textbf{0 1, c}\\\end{array}\right)$
     \\
    \vspace{5ex}
    \\
       $B2_{Face}[:,:] = \left(\begin{array}{l} \textbf{1 0 0}  \\ \textbf{0 1 0 }\\ \textbf{0 0 1} \end{array}\right)$
    \\\\
    This concludes the description of the two-layered generative model.

\end{document}